\documentclass[10pt,twocolumn,aps,prl,superscriptaddress,reprint]{revtex4-2}

\usepackage[latin9]{inputenc}
\usepackage{setspace}

\usepackage{bm}
\usepackage{amsmath}
\usepackage{graphicx}
\usepackage[unicode=true,
 bookmarks=false,
 breaklinks=false,pdfborder={0 0 1},colorlinks=false]
 {hyperref}
\usepackage{breakurl}

\usepackage{float}
\usepackage{threeparttable} 
\usepackage{multirow}
\usepackage{lineno}
\makeatletter
\usepackage{soul,xcolor}
\usepackage{bm}
\usepackage{url}
\usepackage{ulem}
\makeatother
\newcommand{\be}{\begin{equation}}
\newcommand{\ee}{\end{equation}}

\newcommand\redsout{\bgroup\markoverwith{\textcolor{red}{\rule[0.5ex]{2pt}{0.7pt}}}\ULon}

\setstcolor{red}

\begin{document}

\title{ Multigap superconductivity in lithium intercalated bilayer Mo$_2$C  }

\author{Can Hong}
\affiliation{College of Physics \& Optoelectronic Engineering, Department of Physics, Jinan University, Guangzhou 510632, China}	

\author{Danhong Wu}
\affiliation{College of Physics \& Optoelectronic Engineering, Department of Physics, Jinan University, Guangzhou 510632, China}	

\author{Xi-Bo Li}
\affiliation{College of Physics \& Optoelectronic Engineering, Department of Physics, Jinan University, Guangzhou 510632, China}

\author{Feipeng Zheng}
\thanks{Corresponding author}
\email{fpzheng\_phy@email.jnu.edu.cn}
\affiliation{College of Physics \& Optoelectronic Engineering, Department of Physics, Jinan University, Guangzhou 510632, China}

\date{\today}
\begin{abstract}
	Interlayer  coupling can significantly influence the physical properties of layered transition metal compounds. 
	The superconductivity in layered Mo$_2$C systems, belonging to the emergent family of MXene, has garnered considerable attention. 
	However, the impact of interlayer coupling on superconductivity, and the anisotropic superconducting properties in these systems are not yet clear. 
    By performing first-principles calculations of electron-phonon coupling and anisotropic superconducting properties, we show that the interlayer coupling in bilayer 1$T$-Mo$_2$C suppresses superconductivity, resulting in a significant drop in superconducting transition temperature ($T_{\mathrm{c}}$) from 4.2 $K$ in its monolayer form to nearly 0 $K$.
	By introducing lithium atoms into the interlayer space of the bilayer, the interlayer coupling can be effectively weakened, transforming the system into a two-gap superconductor with a $T_{\mathrm{c}}$ above 10 $K$. 
    A 3\% tensile strain can further transform the system into a three-gap superconductor with a significantly enhanced $T_{\mathrm{c}}$ of approximately 24.7 $K$, which is very high in the Mo$_2$C related systems. 
    The enhancement of the superconductivity induced by the strain is mainly due to the downshift of an energy band with a flat dispersion to the energy near the Fermi level.
	The in-plane vibrations of Mo atoms and the $d$-orbital electrons of Mo atoms are most important for the formation of the superconductivity. 
	Our method can also be applied to multilayer Mo$_2$C systems. 
	Given the successful synthesis of layered Mo$_2$C systems and the experimental realization of alkaline metal atom depositions, our work presents a practically feasible strategy for achieving high $T_{\mathrm{c}}$ and multigap superconductivity in layered Mo$_2$C.
\end{abstract}
\maketitle

\section{Introduction}
The interlayer coupling  effect can vary across different layered systems.
Take, for instance, the case of $2H$-NbSe$_2$: the transition from a monolayer to a multilayer results in the enhancement of superconductivity while the weakening of the charge density wave \citep*{Xi2015}. 
In the case of 1$T'$-WTe$_2$, the transition from a monolayer to a multilayer structure results in an enhancement of its metallic properties, causing a decrease in the magnitude of the topological band gap from a positive to a negative value  \citep*{Qian2014a,Zheng2016d}.
While in 1$T$-NiTe$_2$, superconductivity is suppressed on going from a monolayer to a bilayer \citep*{Zheng2019c}. 
Through interlayer coupling manipulation, their physical properties  can be effectively tuned, leading to more diverse phases.
For example, intercalating bulk 2$H$-NbSe$_2$ with cations allows it to retain its high superconducting transition temperature ($T_{\mathrm{c}}$) while exhibiting the Ising superconductivity observed in its monolayer counterpart \citep*{Zhang2022}.
Incorporating hydrate molecules to decouple the interlayer of bulk 2$H$-NbSe$_2$ makes it nearly identical to the behaviors found in NbSe$_2$ monolayers with high charge density wave transition temperature and coexisting Ising superconductivity \citep*{Sun2023}.
Intercalation of transition metal atoms enables the change of charge order \citep*{Wu2023,Luo2023a}, introduction of magnetic order \citep*{Huan2022}, and superconductivity \citep*{pan2022}.
Therefore, studying the interlayer coupling effects in layered compounds not only deepens the understanding of their physical properties, but also offers valuable insights for customizing these properties.

Transition metal chalcogenides have garnered widespread attention due to their rich physical properties, such as the interplay between charge density waves and superconductivity, two-dimensional superconductivity, topologically nontrivial states, and magnetic order. 
More recently, the scope of investigation has extended to emergent transition metal compounds, two-dimensional transition metal nitrides, and carbonitrides, so-called MXenes. 
Among these, Mo$_2$C has received considerable interest, owing to its feasible synthesis, fascinating properties, and promising applications.
Mo$_2$C can be crystalline in multi-structure phases, among which orthorhombic $\alpha$ phase, and hexagonal $H$ and $T$ phases are of particular interest.
They are experimentally feasible \citep*{Xu2015a,Zhao2019} and have been shown to be candidate materials for the study of the interplay between different orders due to the presence of superconductivity \citep*{Xu2015a,Zhao2020f,Zhang2019b,Fan2020,Liu2019,Zhang2021,Ge2021,Lei2017a,Zhang2017b,Bekaert2020,Bekaert2022,Zhao2023a,Wang2023}, the potential nonlinear Hall effect \citep*{Zhao2023a}, and topological nontrivial states \citep*{Zhao2020f}.
Among the above studies, extensive investigations have been conducted on the superconductivity of $\alpha$-Mo$_2$C, encompassing both experimental and theoretical aspects.
In detail, ultrathin $\alpha$-Mo$_2$C has been synthesized and found to exhibit superconductivity with $T_{\mathrm{c}} = 4~K$ \citep*{Xu2015a}. 
The superconductivity in this system has been observed to be weakened  either with decreasing thickness \citep*{Xu2015a}, or in the regions near grain boundaries \citep*{Liu2019}. 
Further research indicates that the superconductivity is robust against defects \citep*{Zhang2019b}.
Subsequent computational study reveals that the hardness of the superconducting $\alpha$-Mo$_2$C is comparable to common alloys \citep*{Ge2021}.

In contrast to the extensive experimental and theoretical investigations of superconductivity in $\alpha$-Mo$_2$C, research on the superconductivity of layered $H$- and $T$-Mo$_2$C is currently predominantly on the theoretical aspect.
Theoretical calculations reveal that monolayer 1$T$-Mo$_2$C and 1$H$-Mo$_2$C both exhibit superconductivity, with $T_{\mathrm{c}}$ approximately 5.9 K and 3.2 K, respectively \citep*{Lei2017a,Zhang2017b}. 
Several approaches have been proposed to enhance their $T_{\mathrm{c}}$s, including hydrogen adsorption \citep*{Lei2017a,Bekaert2022}, halogen atom adsorption \citep*{Zhang2017b}, and biaxial strain \citep*{Liu2023}.
Hydrogen adsorption on the surfaces of monolayer 1$H$-Mo$_2$C and 1$T$-Mo$_2$C leads to an increase in the $T_{\mathrm{c}}$ to approximately 12.6 $K$ \citep*{Lei2017a,Bekaert2020}. 
Additionally, Br atom adsorption enhances the $T_{\mathrm{c}}$ of monolayer 1$T$-Mo$_2$C to 12.8 $K$ \citep*{Zhang2017b}. 
Applying a 4\% tensile strain raises the $T_{\mathrm{c}}$ of monolayer 1$T$-Mo$_2$C to 6.7 $K$, while a 10\% tensile strain increases the $T_{\mathrm{c}}$ of monolayer 1$H$-Mo$_2$C to 11.8 $K$ \citep*{Liu2023}. 
Furthermore, a recent study suggests that the out-of-plane heterostructure formed by a monolayer 1$T$-Mo$_2$C and 1$H$-Mo$_2$C exhibits nonlinear Hall effects and potential Ising superconductivity \citep*{Zhao2023a}.
However, there are several issues regarding the study of layered $H$ and $T$ phases of Mo$_2$C. 
Firstly, the existing studies are mostly focused on monolayer systems. 
The superconductivity in their multilayer counterparts and the effect of interlayer coupling  are unclear. 
Secondly, even when considering a small Coulomb repulsion in the calculations using Coulomb pseudopotential $\mu^{*} = 0.1$, the obtained $T_{\mathrm{c}}$s were mostly below 13 $K$. 
Finally, the previous studies on the superconductivity were mostly using isotropic methods based on McMillan equation  \citep*{McMillan1968,Allen1975} except for the  recent research on the heterostructure \citep*{Zhao2023a}.
Thus, the anisotropic superconducting properties in layered Mo$_2$C, such as the distribution of the superconducting gap and its evolution with temperature, still need investigation, as these quantifies are crucial for the understanding of superconductivity in low-dimensional systems with anisotropic Fermi surfaces~\citep*{Margine2013,Choi2002,Sanna2012}.

In this study, we employed first-principles calculations to investigate the electronic structure, electron-phonon coupling, and anisotropic superconducting properties of few-layer Mo$_2$C, with a particular focus on bilayer configurations. 
We found that bilayer Mo$_2$C exhibits strong interlayer coupling, which acts to suppress its superconductivity, reducing the $T_{\mathrm{c}}$ from 4.2 $K$ in its monolayer counterpart to 0 $K$.
Interestingly, the introduction of lithium atoms into the interlayer space, serving to weaken the interlayer coupling through expanding interlayer space and simultaneously introducing electron doping, transforms the bilayer Mo$_2$C into a two-gap superconductor with a $T_{\mathrm{c}}$ exceeding 10 $K$, surpassing that of its monolayer counterpart. 
Remarkably, applying a small tensile strain of 3\% significantly enhances superconductivity, transforming the system into a three-gap superconductor with a $T_{\mathrm{c}}$ of approximately 24.7 $K$, which is notably high compared to other layered Mo$_2$C-related systems.
The methodologies applied here can also be extended to induce superconductivity in multi-layer Mo$_2$C systems.
Additionally, we will discuss the structure of superconducting gaps and the potential for Ising superconductivity in the system.

\section{Computational Method}

Density-functional theory and density-functional perturbation theory calculations were performed using the PBE exchange-correlation functional \citep*{Perdew1996}, to scrutinize  the crystal structures, electronic structures, and electron-phonon coupling (EPC) of few-layer Mo$_2$C before and after the  alkali metal intercalations \citep*{Kresse1996,Giannozzi2009,Ponce2016}.
The crystal structures were visualized using VESTA package \citep*{Momma2011}.
Nonlocal van der Waals interaction (vdw-df2-b86r) \citep*{Hamada2014,Lee2010a} was taken into account in the calculations of pristine multilayer Mo$_2$C. 
The projector augmented-wave pseudopotentials~\citep*{kresse1999ultrasoft} were used to describe the interaction between the valance and core electrons.
The Kohn-Sham valence states were expanded using plane waves, with energy cutoffs set at 50 Ry and 500 Ry for wave functions and charge densities, respectively. 
An 18$\times$18$\times1$ $\boldsymbol{k}$ mesh and a 6$\times$6$\times$1 $\boldsymbol{q}$ mesh were used to calculate the ground states of charge densities and phonons for bilayer systems, whereupon the EPC matrix elements, $g_{mn,\nu}(\boldsymbol{k},\boldsymbol{q})$, were calculated~\citep*{Ponce2016}.
The matrix elements quantify the scattering amplitude between the electronic states with a wave vector $\boldsymbol{k}$, a band index $m$ [denoted as ($\boldsymbol{k},m)$], and ($\boldsymbol{k}$+$\boldsymbol{q}$, n)  through a phonon mode with a branch $\nu$ and a wave vector $\boldsymbol{q}$.
The above quantities were further interpolated~\citep*{mostofi2008wannier90} to a $120 \times 120 \times 1$ $\boldsymbol{k}$ grid and a $60 \times 60 \times 1$ $\boldsymbol{q}$ grid, whereby the Eliashberg function $\alpha^2F(\omega)$ was calculated.
The Eliashberg function is defined as
$$
\alpha^{2} F(\omega)=\frac{1}{2} \sum_{\nu} \int_{\mathrm{BZ}} \frac{\mathrm{d}\boldsymbol{q}}{\Omega_{\mathrm{BZ}}} \omega_{\boldsymbol{q}\nu} \lambda_{\boldsymbol{q}\nu} \delta\left(\omega-\omega_{\boldsymbol{q}\nu}\right),
$$
where $\lambda_{\boldsymbol{q}\nu}$ is a phonon-momentum-resolved EPC constant, $\Omega_{\mathrm{BZ}}$ the volume of the first Brillouin zone. 
The Dirac delta function $\delta\left(\omega-\omega_{\boldsymbol{q}\nu}\right)$ is approximated by a Gaussian function with a broadening of 0.5 meV.
The $\boldsymbol{k}$-resolved superconducting gaps on the Fermi surface at a temperature $T$, denoted as $\Delta(\boldsymbol{k},T)$, were determined by solving the anisotropic Midgal-Eliashberg equations on an imaginary axis \citep*{Margine2013}, and were subsequently analytically continued to a real axis with Pad\'{e} functions.
In solving the equations, the Kohn-Sham states within 100 meV around the Fermi level were included, and the Matsubara frequencies were cut off at 0.32 eV.
The Coulomb pseudopotential $\mu^{*}$ used for the calculations of superconductivity, was estimated by $\mu^{*} \approx 0.26 N(0) /[1+N(0)]$, where $N(0)$ is the electronic density of states at the Fermi level \citep*{Bennemann1972}.

\section{Results and discussions}

\subsection{Physical Properties of Bilayer 1$T$-Mo$_2$C}
\begin{figure}
	\centering
	\includegraphics[width=76 mm]{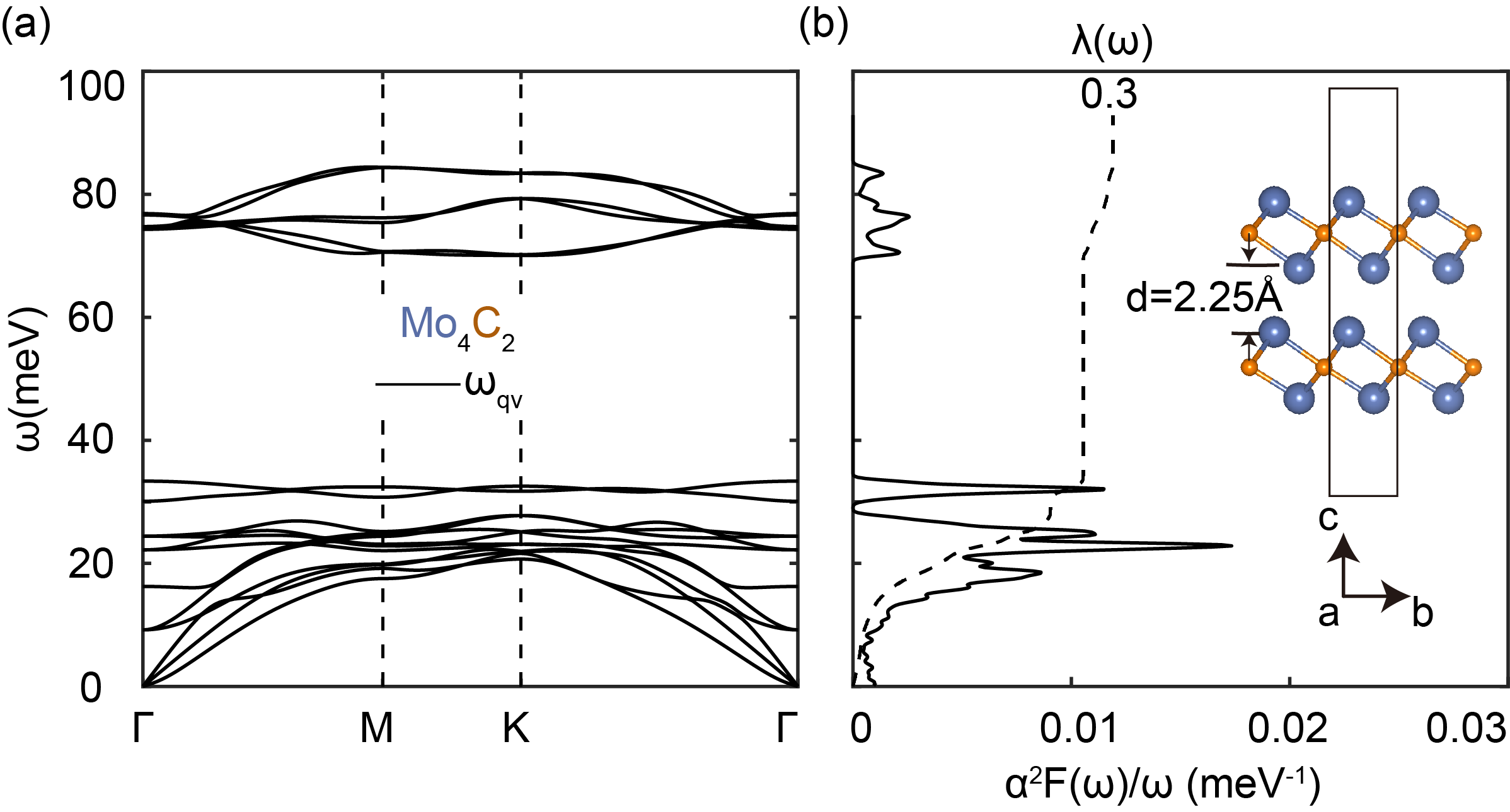} 
		\caption{ The phonon dispersion $\omega_{\boldsymbol{q}\nu}$ (a), and Eliashberg function $\alpha^2F(\omega)$ with the accumulated EPC constant $\lambda(\omega)$ (b) of Mo$_4$C$_2$. 
		The inset in (b) shows the side view of the crystal structure of  Mo$_4$C$_2$.
		The black rectangle represents a unit cell.
		}
	\label{fig1}
\end{figure}

We start by examining the crystal structure, electronic properties, and electron-phonon coupling in bilayer Mo$_2$C. 
Our calculations indicate that bilayer 1$T$-Mo$_2$C (hereinafter referred to as Mo$_4$C$_2$) exhibits the lowest total energy when compared to  $H$ and $\alpha$ phases~(Sec.~S1 \citep{SM}). 
The optimized hexagonal lattice constant for the Mo$_4$C$_2$  is 3.05 \AA, slightly larger than that of its monolayer counterpart.
The Mo$_4$C$_2$ is formed by the $AA$ stacking of two 1$T$-Mo$_2$C monolayers, separated by a van der Waals (vdW) gap of approximately 2.25 \AA, as illustrated in the inset of Fig.~\ref{fig1}(b). 
In each monolayer, every carbon atom occupies the central position within an edge-shared octahedron, formed by six adjacent Mo atoms.
This structure differs from that of transition metal dichalcogenides in the same phase, where the transition metal atoms occupy the center of the octahedron. 
The calculated phonon dispersion [$\omega(\boldsymbol{q})$] depicted in Fig.~\ref{fig1}(a) provides the evidence of the dynamical stability of Mo$_4$C$_2$ crystal, as it reveals the absence of any imaginary phonon modes.
The calculated Eliashberg function [$\alpha^2F(\omega)$] and the integrated EPC [$\lambda(\omega)$] depicted in Fig.~\ref{fig1}(b) show that the total EPC constant ($\lambda$) of the Mo$_4$C$_2$ is only 0.3.
Using the Allen-Dynes-modified McMillan equation \citep*{McMillan1968,Allen1975} with an estimated Coulomb pseudopotential $\mu^{*} = 0.157$, the superconducting $T_{\mathrm{c}}$ is estimated to be 0.0 $K$.
In contrast, the calculated $\lambda$ and $T_{\mathrm{c}}$ of monolayer 1$T$-Mo$_2$C are 0.76 and 4.2 $K$ (Sec.~S2 \citep{SM}), respectively, which are in line with a previous work \citep*{Bekaert2020}.
This clearly demonstrates that EPC suppression occurs in the Mo$_4$C$_{2}$ compared to its monolayer counterpart, implying an adverse effect of interlayer coupling on the EPC.
Additionally, the interlayer distance in  Mo$_4$C$_2$ is only 2.25 \AA~according to our calculation, significantly smaller than that of commonly encountered transition metal compounds like 2$H$-NbSe$_2$ (2.94 \AA) \citep*{Marezio1972}, bilayer $1T$-NiTe$_{2}$ (2.63 \AA) \citep*{Zheng2019c}, and bilayer 1$T$-PtTe$_2$ (2.57 \AA) \citep*{Wu2021}, suggesting a strong interlayer coupling in the Mo$_4$C$_2$.
Indeed, our calculations, as shown in Sec. S3  \citep{SM}, demonstrate that increasing the interlayer spacing in Mo$_4$C$_2$ effectively enhances $N(0)$, thereby providing additional electronic states for EPC.
The impact of interlayer coupling on the $N(0)$ in Mo$_4$C$_2$ resembles that of NiTe$_2$, and is in contrast to NbSe$_2$ and WTe$_{2}$, where metallicity is reinforced in their multilayer counterparts. 
Alkaline metal intercalation provides an effective method to relieve the interlayer coupling by expanding the interlayer space. 
By this method, the interlayer interaction was reduced in twisted bilayer MoS$_2$, enabling fast Li-ion diffusion \citep*{Wu2022a}. 
Furthermore, superconductivity can also be induced in group-IV honeycomb structures \citep*{Flores-Livas2015} and bilayer transition metal dichalcogenides \citep*{Zheng2019c,Wu2021}.
Thus, it is  expected that such method can be applied to  Mo$_4$C$_2$.

\subsection{Method for Preparing Lithium Intercalated Mo$_4$C$_2$ }

\begin{figure}
	\centering
	\includegraphics[width=76 mm]{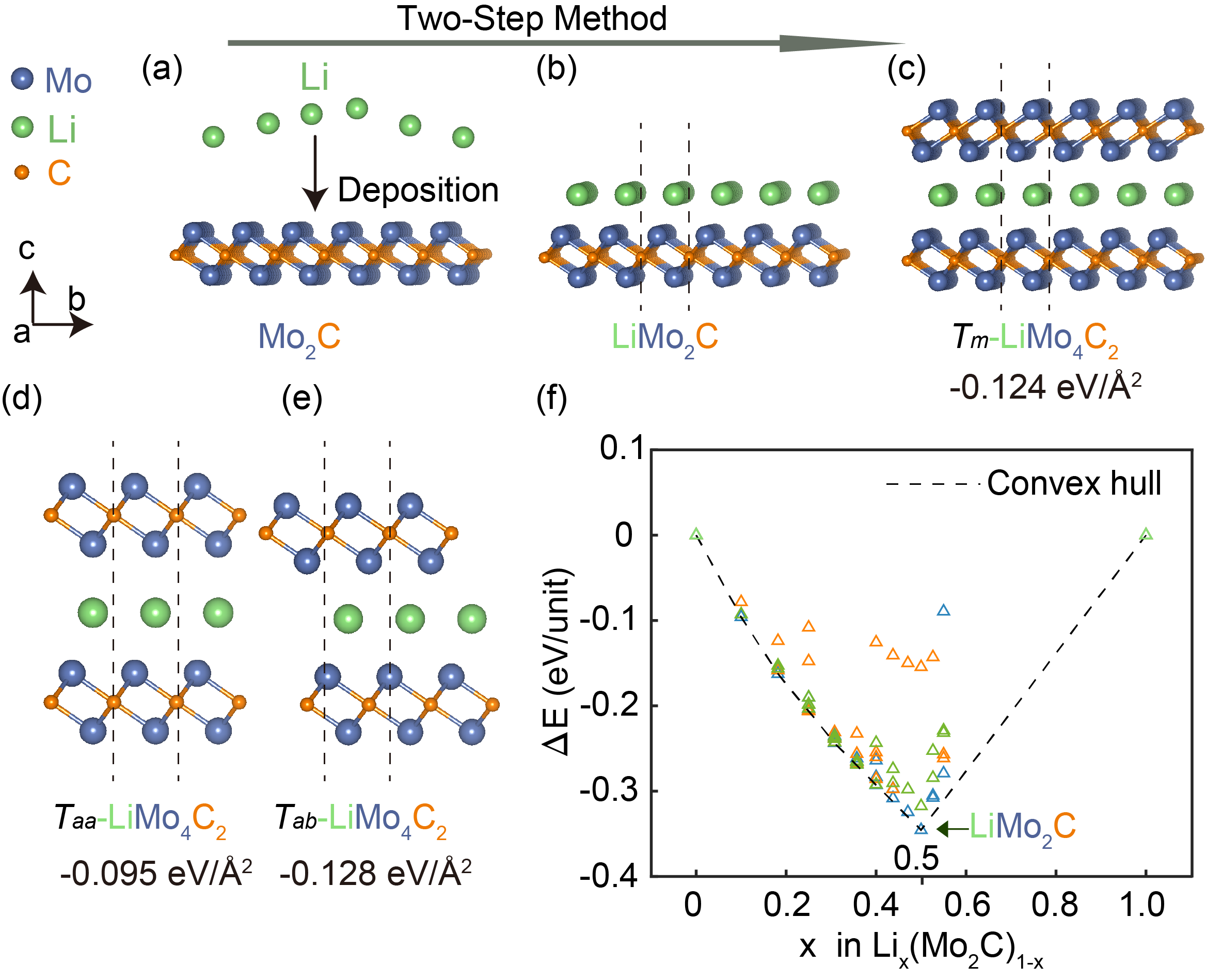} 
		\caption{
	    The proposed two-step process for preparing LiMo$_4$C$_2$. 
		Firstly, lithium atoms are deposited on a monolayer 1$T$-Mo$_2$C surface as shown in  (a) and (b).
		The corresponding formation energies as a function of lithium concentrations ($x$) are presented in (f), along with a convex Hull (see main text for details).
		The blue, orange, and green triangles represent the formation energies of the configurations where the lithium atoms sit above the bottom Mo atoms, top Mo atoms, and C atoms, respectively (see Sec. S5  \citep{SM} for more details).
		Secondly, the  LiMo$_2$C  is further covered by a monolayer Mo$_2$C to form LiMo$_4$C$_2$.
		Three configurations of LiMo$_4$C$_2$  with  formation energies of -0.124, -0.095, and -0.128 eV/\AA$^2$~are displayed in (c)--(e), respectively.
		}
	\label{fig2}
\end{figure}

Our findings suggest that it is energetically unfavorable to directly insert alkaline metal atoms into the interlayer space of Mo$_4$C$_2$, primarily due to strong interlayer coupling (see Sec. S4~ \citep{SM}). 
Alternatively, we identified an energetically favorable method to create alkali-metal intercalated Mo$_4$C$_2$, consisting of two steps as illustrated in Fig.~\ref{fig2}. 
Firstly, lithium atoms were deposited onto the surface of a monolayer 1$T$-Mo$_2$C. 
The chemical formula for a unit of the lithium-adsorbed monolayer Mo$_2$C can be denoted as Li$_m$(Mo$_2$C)$_{n}$, where $m$ and $n$ are integer numbers.
Alternatively, it can be expressed as Li$_x$(Mo$_2$C)$_{1-x}$, with $x = m/(m+n)$ denoting the lithium concentration.
To investigate the thermally stable compositions of Li$_x$(Mo$_2$C)$_{1-x}$, we calculated the formation energies as a function of the $x$, defined as $\Delta E(x)= E[\mathrm{Li}_x(\mathrm{Mo}_2\mathrm{C})_{1-x}] - x E[\mathrm{Li}] - (1-x) E[\mathrm{Mo}_2\mathrm{C}]$, where $E[\mathrm{Li}]$ and $ E[\mathrm{Mo}_2\mathrm{C}]$ represent the energies of a body-centered cubic lithium crystal per atom and a monolayer 1$T$-Mo$_2$C crystal per Mo$_2$C, respectively.
There are three distinct adsorption sites on the 1$T$-Mo$_2$C monolayer surface where lithium atoms can bind, including the sites above the carbon atoms, the sites above the top-layer Mo atoms, and the sites above the bottom-layer Mo atoms [see Figs. S4(b) and S4(d)  \citep{SM}].
The computed results, depicted in Fig.~\ref{fig2}(f), reveal that when all sites above the bottom-layer Mo atoms are occupied (corresponding to $x = 0.5$), the resultant LiMo$_{2}$C [Li$_{0.5}$(Mo$_{2}$C)$_{0.5}$] configurati on [Fig.~\ref{fig2}(b)] exhibits the most negative formation energy and resides on the convex hull, as indicated by the blue triangle at $x = 0.5$.
When additional Li atoms are deposited ($x > 0.5$) and the structures are optimized, the resulting formation energies start to increase, deviating from the convex hull, as illustrated in Fig.~\ref{fig2}(f). 
These findings underscore the thermal stability of the LiMo$_{2}$C with the lithium atoms adsorbed above the bottom Mo atoms, indicating its resistance to decomposition into other compounds with different stoichiometries.
The phonon dispersion calculation shown in Fig.~S5  \citep{SM} further confirms the dynamical stability of the LiMo$_{2}$C.
We also considered the deposition of lithium on a monolayer 1$H$-Mo$_2$C.
However, we found that there can be a phase transition from 1$H$ to 1$T$ when  lithium atoms adsorb on the monolayer 1$H$-Mo$_2$C surface,  reminiscent of the same phase transition in layered MoS$_2$ induced by Li interaction \citep*{Wu2022a} (see Sec.~S5  \citep{SM}).

After analyzing the Li depositions, we found that a monolayer Mo$_2$C is energetically favorable  to  cover the Li adlayer, forming the Li intercalated bilayer LiMo$_{2}$C with a stoichiometry of LiMo$_{4}$C$_{2}$.
We identified three different configurations with the same stoichiometry of LiMo$_{4}$C$_{2}$, but with different  stacking orders of the  two internal  monolayers.
For the $T_{m}$-LiMo$_{4}$C$_{2}$, as illustrated in Fig.~\ref{fig2}(c), the two monolayers are related  by a horizontal mirror symmetry across the Li adlayer.
In the $T_{aa}$-LiMo$_{4}$C$_{2}$ configuration, as shown in Fig.~\ref{fig2}(d), the two monolayers are $AA$ stacked.
For the $T_{ab}$-LiMo$_{4}$C$_{2}$, as depicted in Fig.~\ref{fig2}(e), the two monolayers stack in the order similar to the case of graphite.
By calculating their formation energies defined as $(E[\mathrm{LiMo}_4\mathrm{C}_2] - E[\mathrm{LiMo}_2\mathrm{C}] - E[\mathrm{Mo}_2\mathrm{C}])/S$, where the $S$ donates an area of a two-dimensional unit cell for the $\mathrm{LiMo}_2\mathrm{C}$, we found that the $T_{m}$, $T_{ab}$ and $T_{aa}$ phases have the formation energies of -0.124, -0.128 and -0.095 eV/\AA$^{2}$, respectively, as labeled in Figs.~\ref{fig2}(c)--~\ref{fig2}(e).
The above results suggest that the $T_{m}$-LiMo$_{4}$C$_{2}$ and  $T_{ab}$-LiMo$_{4}$C$_{2}$ have relatively low formation energies and are energetically more favorable to be synthesized. 
The dynamical stabilities of $T_{m}$-LiMo$_{4}$C$_{2}$ and $T_{ab}$-LiMo$_{4}$C$_{2}$ were also confirmed by the phonon calculations as displayed in Figs.~\ref{fig4}(a) and S9(a)  \citep{SM}, respectively.
Given the relative lower formation energies of $T_{m}$-LiMo$_{4}$C$_{2}$ and $T_{ab}$-LiMo$_{4}$C$_{2}$, and the more significant EPC enhancement in $T_{m}$-LiMo$_{4}$C$_{2}$ which we will show below, our focus will be on the $T_{m}$-LiMo$_{4}$C$_{2}$ in the main text.

\begin{figure}
	\centering
	\includegraphics[width=76 mm]{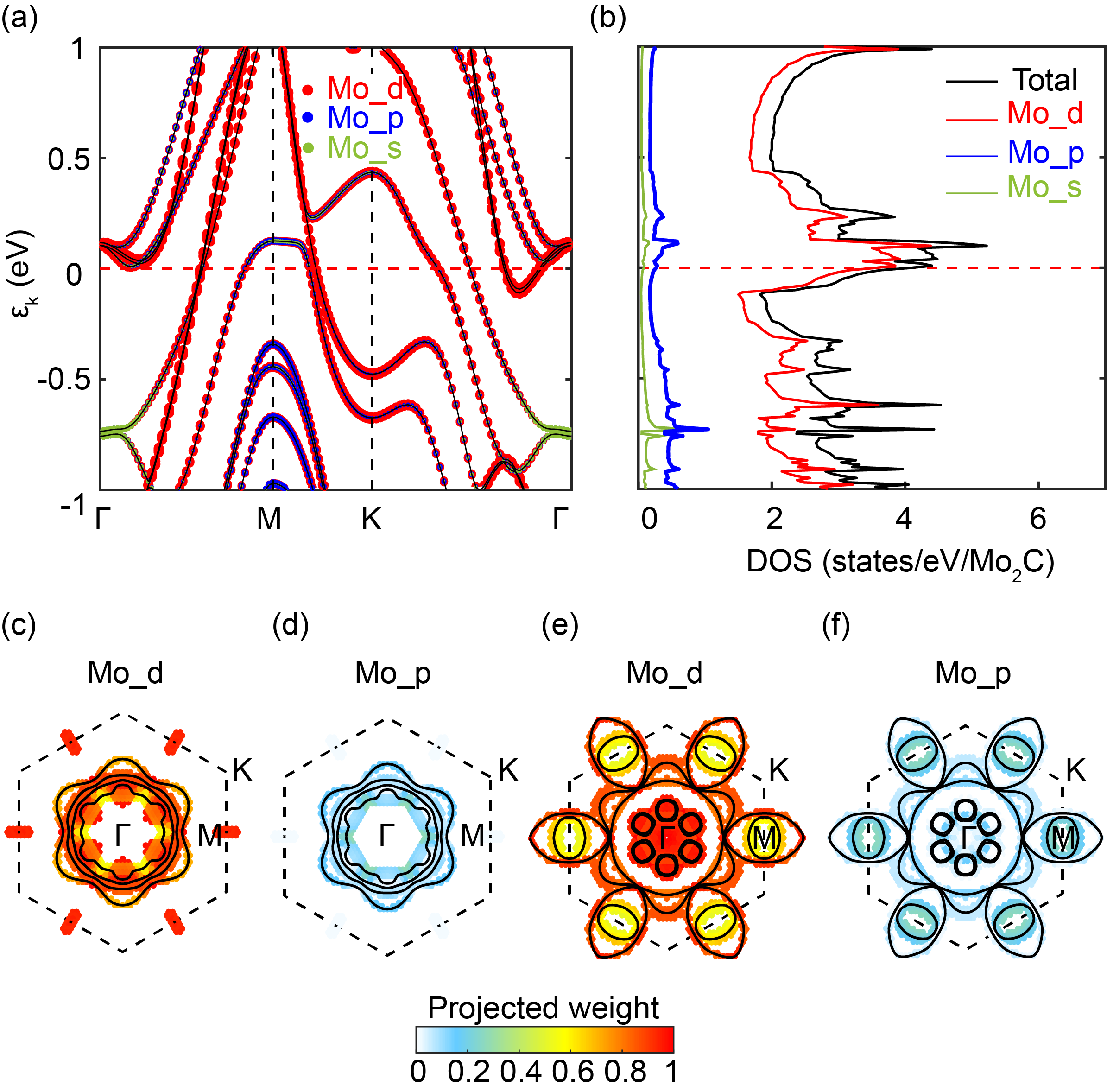} 
		\caption{
		The Projected electronic bandstructure (a), density of states (b), and Fermi surface (e)--(f) onto the atomic orbitals of Mo atoms for $T_{m}$-LiMo$_{4}$C$_{2}$. 
	    (c--d) The projected Fermi surface onto  the atomic orbitals of Mo atoms for Mo$_{4}$C$_{2}$.
		In  (c)--(f), the Fermi surfaces are represented by solid black curves, while the boundaries of the corresponding Brillouin zones are indicated by dashed lines.
		}
	\label{fig3}
\end{figure}

\subsection{Electronic Structure of $T_{m}$-LiMo$_{4}$C$_{2}$ }

We anticipate that the weakening of interlayer coupling in $T_{m}$-LiMo$_{4}$C$_{2}$ will increase $N(0)$, as we have previously analyzed.
Fig.~\ref{fig3}(a) illustrates the electronic bandstructure of $T_{m}$-LiMo$_{4}$C$_{2}$, demonstrating its metallic nature.
Several bands cross the Fermi level, resulting in electron and hole Fermi pockets, as indicated by black circles in Figs.~\ref{fig3}(e) and ~\ref{fig3}(f).
There is a pair of conduction bands across the Fermi level, with a slight energy splitting attributed to weak interlayer coupling. 
Their band minima along the $\boldsymbol{k}$-path of $\Gamma$-$K$ are approximately -0.1 eV below the Fermi level, while those along $\Gamma$-$M$ are slightly above it. 
This results in six electronic pockets near $\Gamma$, as depicted in Figs.~\ref{fig3}(e) and ~\ref{fig3}(f).
By projecting the electronic states near the Fermi level onto atomic orbitals, as shown in Figs.~\ref{fig3}(e) and ~\ref{fig3}(f), it is evident that the electronic states around these six pockets are primarily derived from the Mo-$d$ orbitals. 
Furthermore, a valence band, mainly derived from the Mo-$d$ with slight admixtures of  Mo-$p$ and Mo-$s$ orbitals, intersects the Fermi level around the $M$ points, with its band maximum approximately 0.13 eV above Fermi level [Fig.~\ref{fig3}(a)]. 
This leads to a hole pocket centered at $M$ [Figs.~\ref{fig3}(e)--~\ref{fig3}(f)].
The multiband Fermi surface of $T_{m}$-LiMo$_{4}$C$_{2}$ leads to a calculated $N(0) =  3.87$ states/eV/Mo$_2$C, which is more than twice as large as that of Mo$_{4}$C$_{2}$.
This finding is consistent with the calculated Fermi surface of Mo$_{4}$C$_{2}$, as shown in Figs.~\ref{fig3}(c) and~\ref{fig3}(d), where the electronic states near the Fermi level are clearly fewer compared with those in $T_{m}$-LiMo$_{4}$C$_{2}$.

\subsection{Enhanced Electron-Phonon Coupling in $T_{m}$-LiMo$_{4}$C$_{2}$ }

\begin{figure}
	\centering
	\includegraphics[width=76 mm]{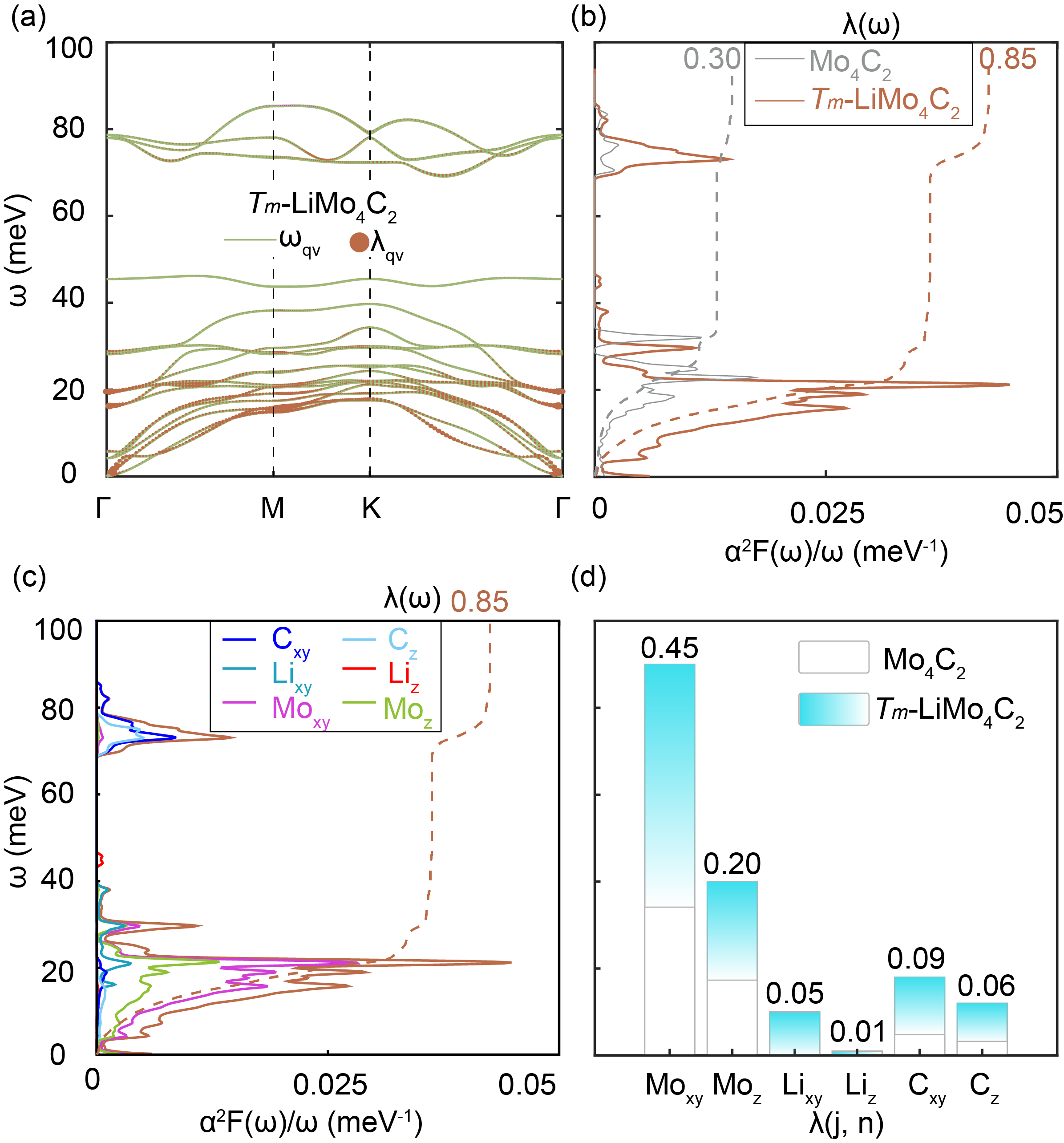} 
		\caption{
		(a) Phonon dispersion ($\omega_{\boldsymbol{q}\nu}$) along with phonon-momentum-resolved EPC constant ($\lambda_{\boldsymbol{q}\nu}$) for $T_{m}$-LiMo$_{4}$C$_{2}$.
		(b) Eliashberg spectra divided by phonon energy [$\alpha^2F(\omega)/\omega$] along with accumulated EPC constant [$\lambda(\omega)$] for  $T_{m}$-LiMo$_{4}$C$_{2}$ and Mo$_{4}$C$_{2}$.
		(c) Projected $\alpha^2F(\omega)/\omega$ onto in-plane ($xy$) and out-of-plane ($z$) vibrations of each atom in $T_{m}$-LiMo$_{4}$C$_{2}$.
		(d) Atomic-vibration-resolved EPC constant for $T_{m}$-LiMo$_{4}$C$_{2}$, obtained by integrating the corresponding spectra in (c).
		The results of Mo$_4$C$_2$ are also shown for comparison.
		}
	\label{fig4}
\end{figure}

The increase of $N(0)$ leads to the increase of EPC in  $T_{m}$-LiMo$_{4}$C$_{2}$.
The calculated $\lambda$ of  $T_{m}$-LiMo$_{4}$C$_{2}$ is 0.85, which is nearly triple the value of Mo$_4$C$_2$, as shown in Fig.~\ref{fig4}(b).
By comparing the  Eliashberg functions $\alpha^{2}F(\omega)$ of the above two materials, it is clear that the promotion of $\lambda$ in $T_{m}$-LiMo$_{4}$C$_{2}$ is  primarily attributed to the phonon energy in the region between 0 and 25 meV, where intensive peaks [Fig.~\ref{fig4}(b)], and large phonon-momentum-resolved EPC constant ($\lambda_{\boldsymbol{q}\nu}$) [Fig.~\ref{fig4}(a)] can be observed.
More specifically, the $\lambda(\omega = 25~\mathrm{meV})$ of $T_{m}$-LiMo$_{4}$C$_{2}$ is 0.66, exhibiting a 247\% increase compared to the value of Mo$_4$C$_2$ (0.19).
By decomposing the  $\alpha^{2}F(\omega)$ into the contributions of the different atomic vibrations, we can quantitatively reveal the origin of the EPC enhancement.
The decomposition can be performed using the following equation \citep*{Wu2021}
$$
\alpha^{2} F(\omega, j, \hat{n})=\frac{1}{2} \sum_{v} \int_{\mathrm{BZ}} \frac{d \mathbf{q}}{\Omega_{\mathrm{BZ}}} \omega_{\mathbf{q} v} \lambda_{\mathbf{q} v} \delta\left(\omega-\omega_{\mathbf{q} v}\right)\left|\hat{n} \cdot e_{\mathbf{q}, v}^{j}\right|^{2},
$$
where $e_{\mathbf{q}, \nu}^{j}$ is the component of the atom $j$ in the polarization vector related to $\omega_{\boldsymbol{q}\nu}$, and $\hat{n}$ is the unit projection direction vector, pointing to the in-plane ($xy$) or out-of-plane ($z$) direction.
The integration of $\alpha^{2} F(\omega, j, \hat{n})$ leads to the atomic-vibration-resolved EPC constant $\lambda(j,\hat{n})$ as  illustrated in Fig.~\ref{fig4}(d).
Comparing the calculated $\lambda(j,\hat{n})$ of $T_{m}$-LiMo$_4$C$_2$ and Mo$_4$C$_2$, it is evident that the $\lambda(j,\hat{n})$ exhibits an overall enhancement in $T_{m}$-LiMo$_4$C$_2$.
Notably, the contributions from the in-plane and out-of-plane vibrations of Mo atoms (Mo-$xy$ and Mo-$z$) increase by 153\% and 121\%, respectively, reaching 0.45 and 0.20.
The enhancements of EPC related to Mo-$xy$ and Mo-$z$ vibrations are evident in the $\alpha^{2} F(\omega, j, \hat{n})$ depicted in Fig.~\ref{fig4}(c), where the peaks associated with  Mo-$xy$ and Mo-$z$ are observed within the  energy range of 0 to 25 meV.
The above discussion indicates that the significant EPC enhancement in $T_{m}$-LiMo$_{4}$C$_{2}$ arises from the contributions of low-energy phonons associated with Mo vibrations.

\subsection{Two-gap superconductivity in $T_{m}$-LiMo$_{4}$C$_{2}$}

\begin{figure}
	\centering
	\includegraphics[width=76 mm]{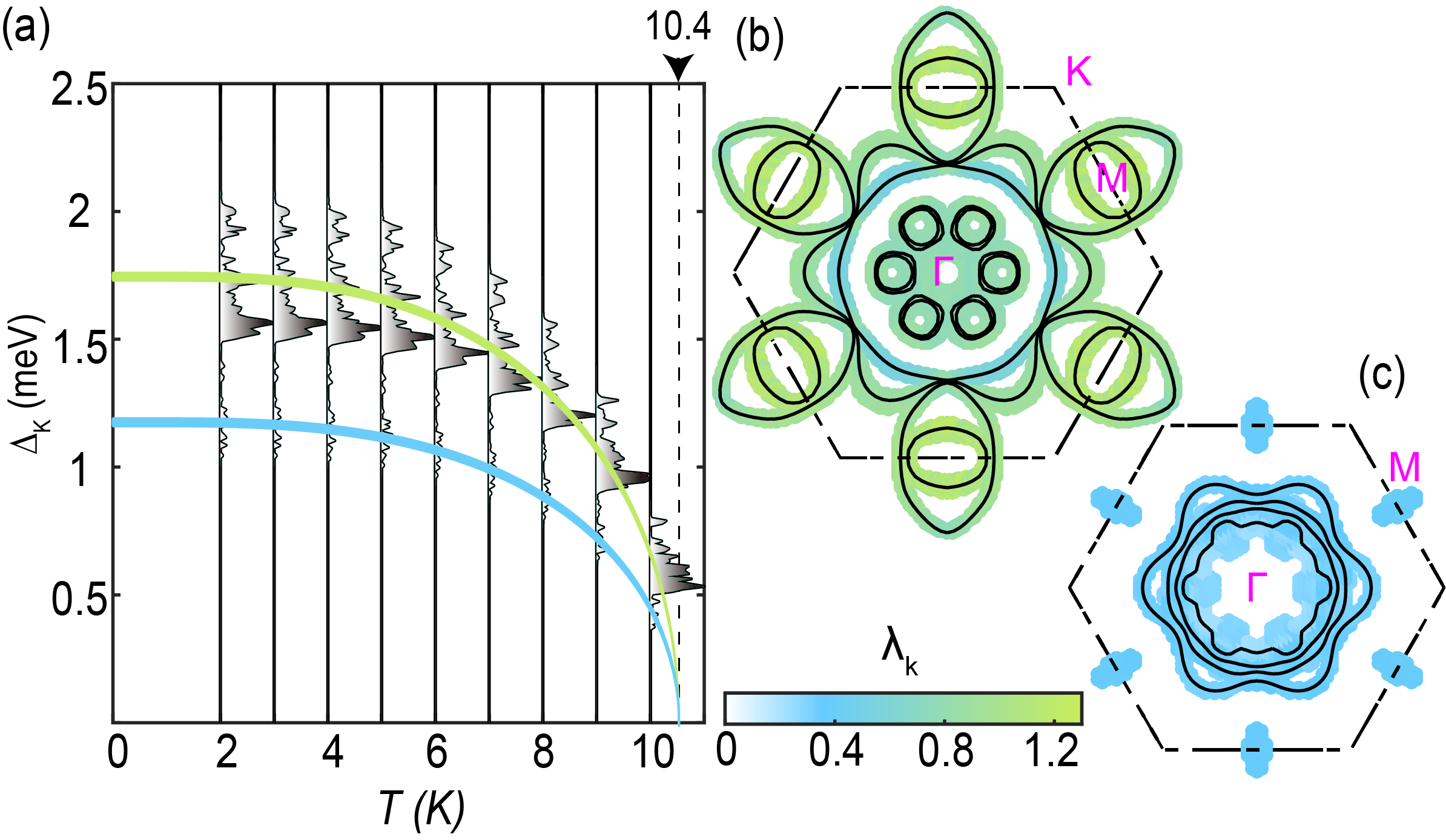} 
		\caption{
		(a) Histograms of $\Delta(\boldsymbol{k},T)$ for $T_{m}$-LiMo$_{4}$C$_{2}$ at various temperatures. 
		The Green and blue curves are the BCS fits of the two gaps. 
		Distribution of $\boldsymbol{k}$-resolved EPC constants near Fermi surface for $T_{m}$-LiMo$_{4}$C$_{2}$ (b) and Mo$_4$C$_2$ (c).
		In (b) and (c), the Fermi surfaces are represented by solid black curves, while the boundaries of the corresponding Brillouin zones are indicated by dashed lines.
		}
	\label{fig5}
\end{figure}

The enhancement of EPC strength and the presence of a multiband Fermi surface result in two-gap superconductivity with an enhanced superconducting $T_{\mathrm{c}}$ in $T_{m}$-LiMo$_{4}$C$_{2}$, as demonstrated below.
We  employ  the imaginary-time anisotropic Migdal-Eliashberg formalism, followed by analytic continuation to the real axis using Pad$\acute{\text{e}}$ functions, through which the $\boldsymbol{k}$-resolved superconducting gaps at Fermi surface under finite temperature, $\Delta(\boldsymbol{k},T)$, can be obtained.
In solving the anisotropic Migdal-Eliashberg equations, the estimated $\mu^{*} = 0.208$ based on $N(0)$ is used (see Sec. II for details) .
The histogram in Fig.~\ref{fig5}(a) illustrates the temperature-dependent variation in the energy distribution of $\Delta(\boldsymbol{k},T)$.
It is evident that the magnitude of $\Delta(\boldsymbol{k},T)$ decreases as the temperature increases.
The $\Delta(\boldsymbol{k},T)$  eventually reaches zero at $T = 10.4~K$, indicating that the calculated  $T_{\mathrm{c}}$ for $T_{m}$-LiMo$_{4}$C$_{2}$ is 10.4$~K$.
Moreover, upon analyzing the structure of $\Delta(\boldsymbol{k},T)$, it is  evident that the system exhibits a two-gap characteristic. 
At a given temperature, most of the $\boldsymbol{k}$ states  exhibit a  higher magnitude of $\Delta(\boldsymbol{k},T)$, while the remaining states display a lower magnitude.
For instance, at $T = 2~K$, the majority of $\boldsymbol{k}$ states exhibit the values of $\Delta(\boldsymbol{k},T)$ ranging from about 1.5 to 2.0 meV, while the remaining $\boldsymbol{k}$ states fall within the range of approximately 1.0 to 1.4 meV.
This can be also confirmed from the distribution of $\Delta(\boldsymbol{k},T = 2~K)$ in the Brillouin zone, as shown in Fig.~S7(b)  \citep{SM}.
The two-gap characteristic is consistent with the distributions of the $\boldsymbol{k}$-resolved EPC constant $\lambda_{\boldsymbol{k}}$ shown in Fig.~\ref{fig5}(b), where the $\boldsymbol{k}$ states around the large circular Fermi pocket at $\Gamma$ have  lower $\lambda_{\boldsymbol{k}}$  values of around 0.4, whereas the other  $\boldsymbol{k}$ states have  higher  $\lambda_{\boldsymbol{k}}$ values close to 1.0. 
The aforementioned circular Fermi pocket indicates isotropic characteristics, resembling the behavior of a free electron gas system.
The enhanced EPC in $T_{m}$-LiMo$_{4}$C$_{2}$ compared to Mo$_4$C$_2$ is also evident by comparing Figs.~\ref{fig5}(b) and ~\ref{fig5}(c), where the $\lambda_{\boldsymbol{k}}$ of $T_{m}$-LiMo$_{4}$C$_{2}$ [Fig.~\ref{fig5}(b)] shows an overall increase.

\begin{figure}
	\centering
	\includegraphics[width=76 mm]{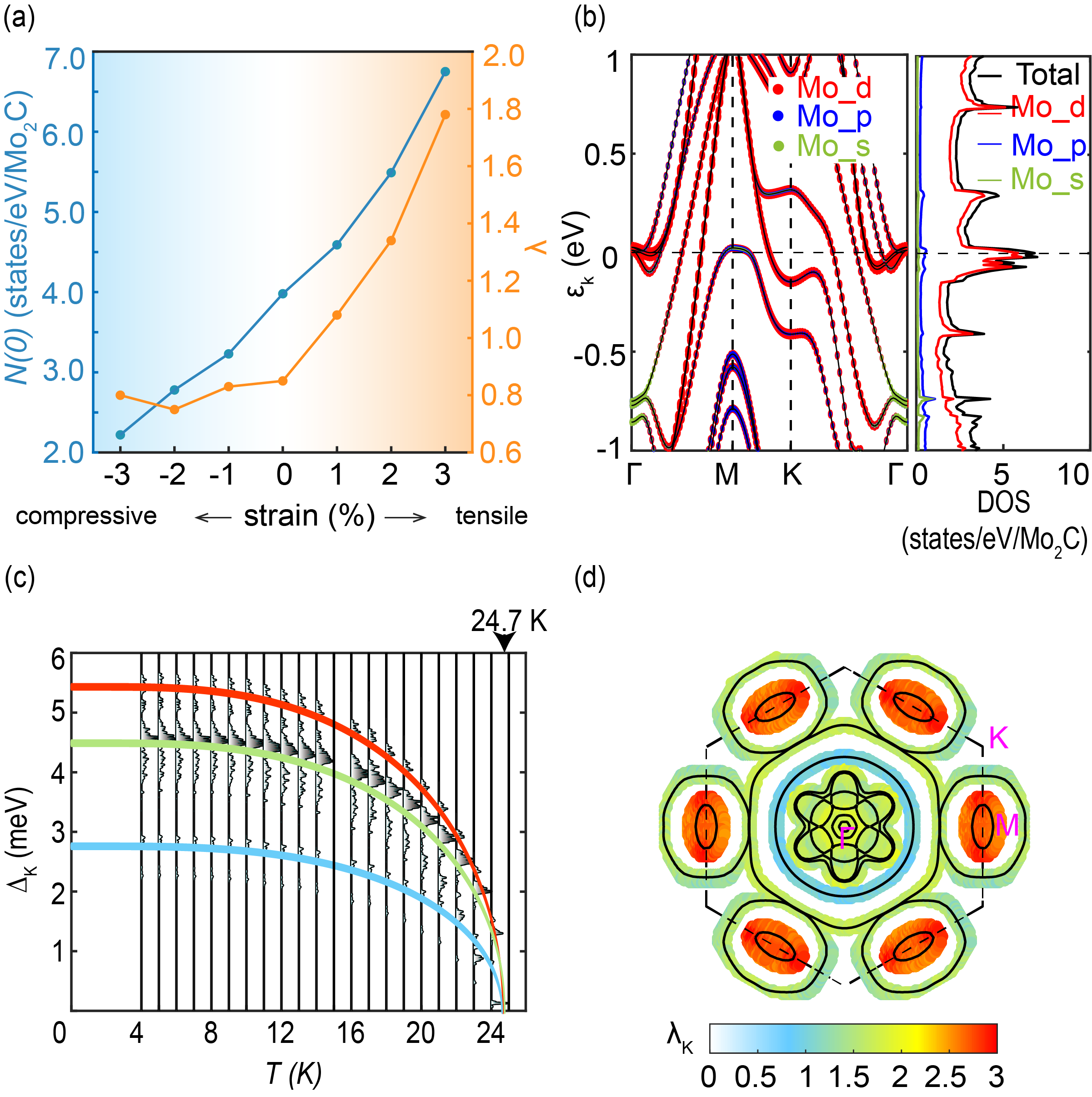} 
	\caption{
	(a) Evolution of $N(0)$ and $\lambda$ as a function of strains for $T_{m}$-LiMo$_{4}$C$_{2}$.
	(b) Projected bandstructure and DOS onto the atomic orbitals for $T_{m}$-LiMo$_{4}$C$_{2}$ under a 3\% tensile strain.
	(c) Histograms of $\Delta(\boldsymbol{k},T)$, and (d) the distribution of $\lambda_{\boldsymbol{k}}$ near Fermi surface for the $T_{m}$-LiMo$_{4}$C$_{2}$ under a 3\% tensile strain.
	In (d), the Fermi surface is represented by solid lack curves, while the boundary of the  Brillouin zone is indicated by dashed lines.
	}
	\label{fig6}
\end{figure}

\subsection{Three-gap superconductivity with high $T_{\mathrm{c}}$ in $T_{m}$-LiMo$_{4}$C$_{2}$ induced by a small tensile strain}

\begin{figure}
	\centering
	\includegraphics[width=76 mm]{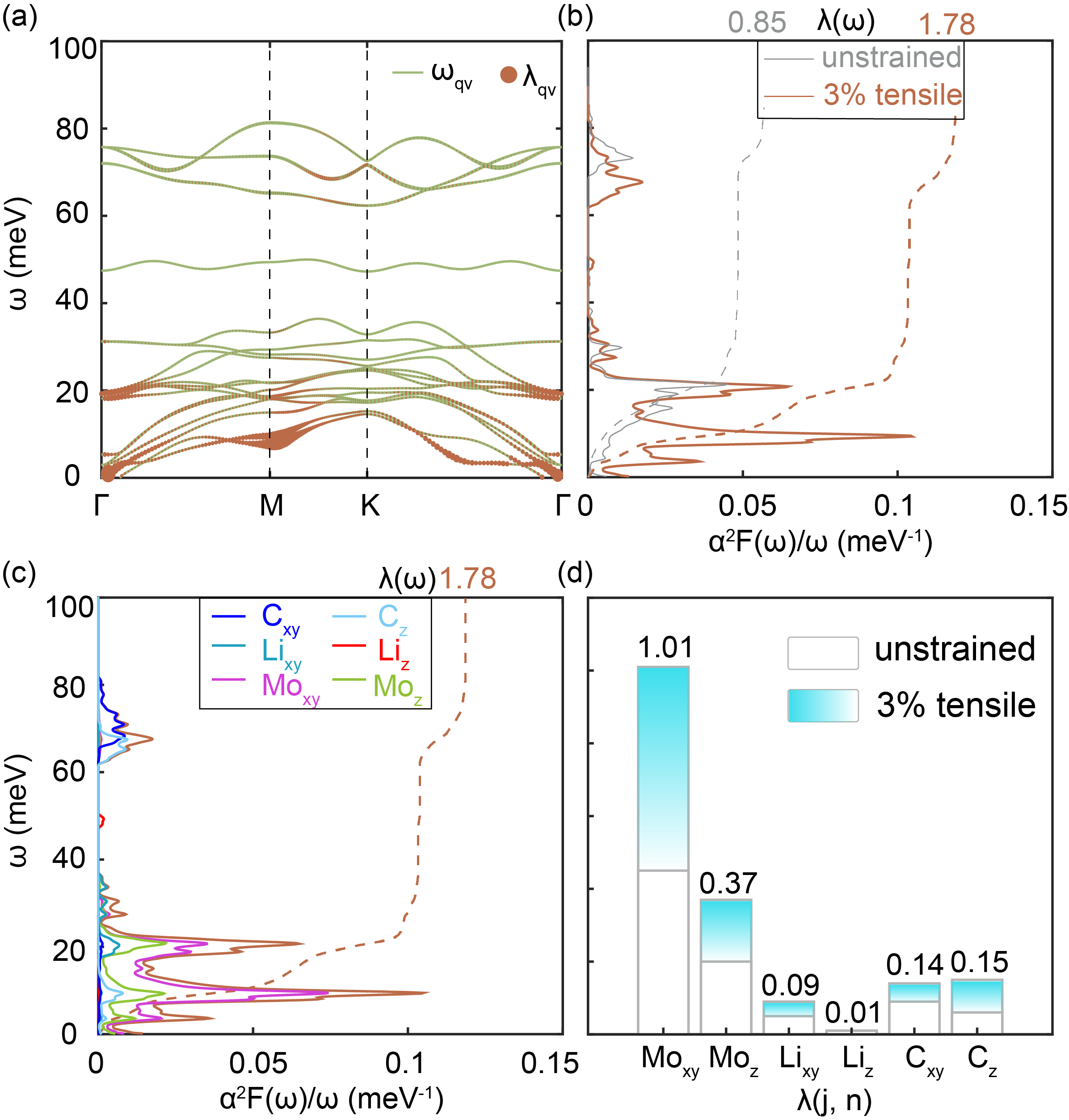} 
		\caption{
        (a) The $\omega_{\boldsymbol{q}\nu}$ of  $T_{m}$-LiMo$_{4}$C$_{2}$ under a 3\% tensile strain, along with the corresponding $\lambda_{\boldsymbol{q}\nu}$.
		(b) The $\alpha^2F(\omega)/\omega$ along with $\lambda(\omega)$ for the  $T_{m}$-LiMo$_{4}$C$_{2}$ with and without the tensile strain.
		(c) Projected $\alpha^2F(\omega)/\omega$ onto in-plane ($xy$) and out-of-plane ($z$) vibrations of each atom in the tensile strained $T_{m}$-LiMo$_{4}$C$_{2}$.
		(d) Atomic-vibration-resolved EPC constant for the tensile strained $T_{m}$-LiMo$_{4}$C$_{2}$, obtained by integrating the corresponding spectra in (c).
		The results of the unstrained  $T_{m}$-LiMo$_{4}$C$_{2}$  are also shown for comparison.
		}
	\label{fig7}
\end{figure}

Given possible in-plane strains in ultrathin materials, which can be induced by a substrate,  the potential discrepancies in lattice constants between PBE calculations and experimental measurements,and the significance contribution to the EPC of $T_{m}$-LiMo$_{4}$C$_{2}$ from Mo$_{xy}$ vibration, it is necessary  to investigate the evolution of EPC and superconductivity in response to in-plane strains for this material.
We apply biaxial strains to $T_{m}$-LiMo$_{4}$C$_{2}$ by adjusting the in-plane lattice constant and then optimizing its internal coordinates.
We define a  biaxial  strain as $(a-a_0)/a_0$, where $a_0 = 3.01$~\AA~is the optimized hexagonal lattice constant of $T_{m}$-LiMo$_{4}$C$_{2}$ without strains,  and $a$ is the lattice constant under strains. 
In Fig.~\ref{fig6}(a), the evolutions of $N(0)$ and $\lambda$ with strains are depicted.
It is evident that both $N(0)$ and $\lambda$ increase with the tensile strain, peaking at 3\% strain with values of approximately 6.8 states/eV/Mo$_2$C and 1.78, respectively.
It is evident from Fig.~\ref{fig7}(a) that the crystal structure of $T_{m}$-LiMo$_{4}$C$_{2}$ under the 3\% strain is dynamically stable.
Further increase the tensile strain leads to the dynamical instability of the $T_{m}$-LiMo$_{4}$C$_{2}$ crystal.
The rise in $N(0)$ under the 3\% tensile strain can be understood by comparing the bandstructure and Fermi surface with strain [Figs.~\ref{fig6}(b) and ~\ref{fig6}(d)] to those without strain [Figs.~\ref{fig3}(a) and ~\ref{fig5}(b)].
The most significant change in the bandstructure after applying the strain is the downward shift of the valence band top near the M points, as evident when comparing Fig.~\ref{fig6}(b) with Fig.~\ref{fig3}(a).
Since the  valence band curvature is relatively flat near its top, this downward shift clearly increases number of the states near the Fermi level.
This is  seen by comparing the DOS depicted in Fig.~\ref{fig3}(b) and ~\ref{fig6}(b), where the peak at 0.13 eV [Fig.~\ref{fig3}(b)] has shifted downward to an energy very close to the Fermi level [Fig.~\ref{fig6}(b)], in accordance with the shift of the valence band top.
Furthermore, the conduction bands, with their bottoms located near $\Gamma$, also undergo downward shifts. 
This results in the merging of six Fermi circles near $\Gamma$ into larger circles centered at $\Gamma$, accompanied by the appearance of additional small Fermi circles around $\Gamma$. 
These alterations in the bandstructure contribute significantly to the increased value of $N(0)$. 
This, in turn, provides a greater number of scattering channels and electronic states available for EPC, which are responsible for the substantial enhancement of $\lambda$ from 0.85 to 1.78 [Fig.~\ref{fig7}(b)].
The superconductivity in $T_{m}$-LiMo$_{4}$C$_{2}$ is expected to be strengthened due to the increased $\lambda$. 
The $\Delta(\boldsymbol{k},T)$ of the strained $T_{m}$-LiMo$_{4}$C$_{2}$ was calculated based on the aforementioned anisotropic Migdal-Eliashberg formalism, with an estimated $\mu^{*} = 0.226$.
The evolution of the $\Delta(\boldsymbol{k},T)$ histogram, as depicted in Fig.~\ref{fig6}(c), clearly indicates that the $T_{\mathrm{c}}$ for  $T_{m}$-LiMo$_{4}$C$_{2}$ under 3\% tensile strain is 24.7 $K$, more than twice the $T_{\mathrm{c}}$ in the strain-free case (10.4~$K$).
Interestingly, a transition from a two-gap superconductor to a three-gap superconductor is clearly discernible when examining the structure of the $\Delta(\boldsymbol{k},T)$ histogram at temperatures below $T_{\mathrm{c}}$, as illustrated in Fig.~\ref{fig6}(c).
For example, at $T = 4~K$, the $\Delta(\boldsymbol{k},T)$ histogram exhibits three primary peaks centered at approximately 5.4, 4.5, and 2.8 meV, respectively.
The distributions of $\lambda_{\boldsymbol{k}}$, as shown in Fig.~\ref{fig6}(d), indicate that the $\boldsymbol{k}$ states with relatively large $\Delta(\boldsymbol{k})$ around 5.4 meV are located around the M points in the Brillouin zone.
The $\boldsymbol{k}$ states with relatively small $\Delta(\boldsymbol{k})$ around 2.8 meV primarily originate from the circular Fermi circles centered at $\Gamma$, akin to the strain-free case discussed previously.
The remaining $\boldsymbol{k}$ states, as shown in Fig.~\ref{fig6}(d), exhibit a moderate magnitude of $\Delta(\boldsymbol{k})$ around 4.5 meV.
By comparing the $\Delta(\boldsymbol{k},T)$ distributions shown in Figs.~\ref{fig5}(b) and ~\ref{fig6}(d), it is clear that the transition of the state from the two gaps to the three gaps is  primarily due to the presence of the $\boldsymbol{k}$ states with large $\lambda_{\boldsymbol{k}}$ values near the M points.
These emergent  $\boldsymbol{k}$ states originate from the valence band top near  the Fermi level, characterized by a notably flat dispersion, as previously mentioned.
The significant variation  of $\lambda_{\boldsymbol{k}}$ for the $\boldsymbol{k}$ states near different Fermi pockets strongly implies an anisotropic Fermi surface. 
This underscores the significance of accounting for the anisotropic behavior of $\Delta(\boldsymbol{k},T)$ when calculating the superconducting properties of the system.

To further elucidate the mechanism of the strain effect on the EPC, we present the $\alpha^2F(\omega)$ plots for both the unstrained and 3\% tensile-strained $T_{m}$-LiMo$_{4}$C$_{2}$  in Fig.~\ref{fig7}(b).
It is evident that the increase in EPC after the strain  primarily stems from the enhancement of EPC in the low-energy phonon range, approximately 0 to 25 meV.
More specifically, the deviation of $\lambda(\omega = 25~\text{meV})$ between these two cases is 0.82, accounting for nearly 88\% of the discrepancy in their $\lambda$ values (0.93).
Examining the $\alpha^{2} F(\omega, j, \hat{n})$ plot presented in Fig.~\ref{fig7}(c) clarifies that the rise in $\lambda$ within the low-energy range mainly results from the enhanced EPC  associated with Mo$_{xy}$. 
This is evident because the $\alpha^{2} F(\omega, \text{Mo}, xy)$ within the energy range of 0 to 25 meV closely matches the $\alpha^2F(\omega)$. 
The computed $\lambda(\text{Mo},xy)$ values for $T_{m}$-LiMo$_{4}$C$_{2}$, before and after the 3\% tensile strain, as depicted in Fig.~\ref{fig7}(d), provide further support for this observation. 
As the $\lambda(\text{Mo},xy)$ value increases substantially by approximately 125\%, rising from 0.45 to 1.01.

The $T_{ab}$-LiMo$_{4}$C$_{2}$ exhibits analogous properties to $T_{m}$-LiMo$_{4}$C$_{2}$.
The unstrained $T_{ab}$-LiMo$_{4}$C$_{2}$ was calculated to be a two-gap superconductor with a $T_{\mathrm{c}}$ of $11.2~K$ [Fig.~S6(a) \citep*{SM}], slightly higher than that of $T_{m}$-LiMo$_{4}$C$_{2}$.
More details can be found in Sec.~S7 \citep{SM}.
Under a 3\% tensile strain, $T_{ab}$-LiMo$_{4}$C$_{2}$ undergoes a similar transition to $T_{m}$-LiMo$_{4}$C$_{2}$, shifting from a two-gap superconductor to a three-gap one with a $T_{\mathrm{c}}$ of $18.9~K$ [Fig.~S11(a)  \citep{SM}], slightly lower than that of $T_{m}$-LiMo$_{4}$C$_{2}$. 
More details can be found in Sec.~S8  \citep{SM}.

We note that the $T_{ab}$-LiMo$_{4}$C$_{2}$  crystal carries inversion symmetry, with inversion centers positioned at the lithium atom sites. 
In contrast, the $T_{m}$-LiMo$_{4}$C$_{2}$ crystal lacks inversion symmetry. 
Noncentrosymmetric two-dimensional superconductors have the potential to host Ising superconductivity, in which their magnitude of upper critical magnetic fields exceed far beyond the Pauli limits due to locking of spins along the out-of-plane direction, stemming from strong spin-orbit coupling.
Recently, it was theoretically predicted that the out-of-plane heterojunction formed by monolayer 1$H$-Mo$_2$C and 1$T$-Mo$_2$C could be an Ising superconductor with a spin-orbit splitting magnitude comparable to that of the gated MoS$_2$ \citep*{Zhao2023a}, a classic Ising superconductor \citep*{saito2016superconductivity}. 
In contrast to the heterojunction, our calculations for the noncentrosymmetric $T_{m}$-LiMo$_{4}$C$_{2}$ suggest that spin-orbit coupling has a minimal impact on the Fermi surface (see Sec.~S9  \citep{SM}), indicating that $T_{m}$-LiMo$_{4}$C$_{2}$ is unlikely to host Ising superconductivity.

In addition to the lithium atoms, we have also explored the utilization of other alkali metal atoms for the intercalation, including sodium, potassium, and rubidium atoms.
We found that only the sodium atoms are energetically favorable to adsorb on the monolayer Mo$_2$C surface. 
The formation energies are also negative when another monolayer Mo$_2$C is covered onto the sodium adlayer to form NaMo$_4$C$_2$.
This suggests that the sodium intercalated bilayer Mo$_2$C can also be prepared (Sec. 10 \citep{SM}). 

Furthermore, in Sec.~S11  \citep{SM}, we show that the lithium intercalations can also increase the $N(0)$ in multilayer Mo$_2$C including trilayer, four-layer, and bulk Mo$_2$C.
The extents of $N(0)$ enhancements due to lithium intercalation in the trilayer, four-layer and bulk Mo$_2$C are similar to that observed in the bilayer, as discussed above.
This suggests that  superconductivity can also be induced by lithium intercalation in  multilayer Mo$_2$C other than the bilayer.

\section{Conclusion}

In summary, we have computationally studied the crystal structure, electronic structure, phonon, EPC, and superconductivity in bilayer Mo$_2$C and related systems. 
We find that strong interlayer coupling diminishes the $N(0)$ of bilayer 1$T$-Mo$_2$C, subsequently reducing the superconducting $T_{\mathrm{c}}$ from 4.2 $K$ in its monolayer counterpart to 0 $K$. 
A two-step method, involving the deposition of lithium atoms and subsequent covered with a Mo$_2$C monolayer, enables the intercalation of lithium atoms, resulting in the formation of $T_{m}$-LiMo$_4$C$_2$ and $T_{ab}$-LiMo$_4$C$_2$ crystals. 
The insertion of lithium atoms not only increases the interlayer spacing of the bilayer system, weakening the interlayer coupling, but also introduces electron doping. 
These combined effects significantly enhance the $N(0)$ and the EPC in the LiMo$_4$C$_2$ systems. 
By computing the anisotropic superconducting properties, we find that both $T_{m}$-LiMo$_4$C$_2$ and $T_{ab}$-LiMo$_4$C$_2$ systems exhibit two-gap superconductivity, with $T_{\mathrm{c}}$s of 10.2 and 11.2 $K$, respectively. 
The vibrational modes of Mo atoms and the anisotropic EPC of the electronic states near the Fermi surface, particularly involving the $d$-orbital states of Mo atoms, play  dominant roles in the formation of the two-gap superconductivity. 
A 3\% biaxial tensile strain can further induce a transition from two-gap to three-gap superconductivity in $T_{m}$-LiMo$_4$C$_2$, with a significantly enhanced $T_{\mathrm{c}}$ of 24.7 $K$. 
This transition is primarily attributed to the downward shift of a flat dispersion near M points to the vicinity of the Fermi level, which not only enhances the $N(0)$, but also brings in the $\boldsymbol{k}$ states with exceptionally strong EPC strengths. 
From the perspective of atomic vibrations, the substantial enhancement of EPC due to the in-plane vibrations of Mo atoms plays a crucial role in this superconducting transition. 
Meanwhile, the 3\% biaxial tensile strain can also induce a  three-gap superconductivity in $T_{ab}$-LiMo$_4$C$_2$, raising the $T_{\mathrm{c}}$ to 18.9 $K$. 
The methods described in this work can be possibly applied to multilayer Mo$_2$C systems. 
Considering that layered Mo$_2$C systems have been successfully synthesized \citep*{Zhao2019}, and the depositions of alkaline metal atoms can also be realized experimentally \citep*{Jaouen2023,Rosenzweig2020a}, our work provides  an experimentally feasible strategy for achieving multigap superconductivity with high transition temperatures in layered Mo$_2$C systems.

\begin{acknowledgements} 
This work is supported by National Natural Science Foundation of China 11804118, Guangdong Basic and Applied Basic Research Foundation (Grant No.2021A1515010041), and the Science and Technology Planning Project of Guangzhou (Grant No. 202201010222). 
The Calculations were performed on  high-performance computation cluster of Jinan University, and Tianhe Supercomputer System.
\end{acknowledgements}

\bibliographystyle{apsrev4-2}

%

\end{document}